\documentclass[10pt]{article}
\usepackage{amsmath,amsfonts,amssymb,theorem,verbatim}

\usepackage{epsfig}

\newtheorem{Thm}{Theorem}

\newtheorem{Lemma}{Lemma}
\newtheorem{Prop}{Proposition}

\newtheorem{Def}{Definition}

\def\BEN{\begin{enumerate}}  \def\BI{\begin{itemize}}
\def\EEN{\end{enumerate}}   \def\EI{\end{itemize}}

\def\nn{\nonumber}
\def\beq{\begin{eqnarray}} \def\eeq{\end{eqnarray}}

\def\eqn#1{\begin{equation}#1\end{equation}}

\def\al*#1{\begin{align*}#1\end{align*}}

\def\ga*#1{\begin{gather*}#1\end{gather*}}

\def\alat*#1#2{\begin{alignat*}{#1}#2\end{alignat*}}
\def\bea{\begin{eqnarray*}}
\def\eea{\end{eqnarray*}}
\def\ml*#1{\begin{multline*}#1\end{multline*}}

 \def\mbf{\mathbf} 
  
\newcommand{\Bf}[1]{{\mbox{\scriptsize\boldmath$#1$}}}

\def\P{{\mathbb P}}
\def\E{{\mathbb E}}  \def\R{{\mathbb R}}

\def\mc{\mathcal}

\def\le{\left} \def\ri{\right} \def\i{\infty}

\def\te#1{\mathrm{e}^{#1}}  \def\td{\text{\rm d}}
 
\def\I{\int} 

\def\T{\tilde}  \def\H{\hat}
\def\WH{\widehat} \def\WT{\widetilde}

\def\a{\alpha} \def\b{\beta}
\def\g{\gamma}     \def\th{\theta}

\def\e{\epsilon} \def\k{\kappa} \def\l{\lambda} \def\m{\mu} 
\def\x{\xi}  \def\nn{\nonumber}   \def\s{\sigma}
\def\t{\tau}    \def\x{\xi} 
  \def\q{\qquad} \def\D{\Delta}

\def\proof{\noindent{\it Proof:\,}}

\newcommand{\exit}{{\mbox{\, \vspace{3mm}}} \hfill\mbox{$\square$}}

\begin{document}
\title{\Large On perpetual American put valuation and first-passage in a regime-switching model with
jumps\footnote{Research supported by the Nuffield Foundation,
grant NAL/00761/G, and EPSRC grant EP/D039053/1.}}

\author{Z. Jiang\thanks{King's College London,
Department of Mathematics, Strand, London WC2R 2LS, UK
\newline email: \texttt{zjunjiang@yahoo.com.cn}
\newline {\em Present address:} School of Finance, Nanjing University of
Finance and Economics, Nanjing, 210046, China} \and M. R.
Pistorius\thanks{King's College London, Department of Mathematics,
Strand, London WC2R 2LS, UK\newline email:
\texttt{martijn.pistorius@kcl.ac.uk}}}


\date{\large\sl In memory of Yumin Jiang}

\maketitle

\begin{abstract}
\noindent In this paper we consider the problem of pricing a
perpetual American put option in an exponential regime-switching
L\'{e}vy model. For the case of the (dense) class of phase-type
jumps and finitely many regimes we derive an explicit expression
for the value function. The solution of the corresponding first
passage problem under a state-dependent level rests on a path
transformation and a new matrix Wiener-Hopf factorization result
for this class of processes.
\bigskip

\noindent{\bf Keywords} American put option $\cdot$ matrix
Wiener-Hopf factorization $\cdot$ phase-type $\cdot$
regime-switching $\cdot$ first-passage problem
\medskip

\noindent{\bf Mathematics Subject Classification (2000)} 60K15
$\cdot$ 90A09
\medskip

\noindent{\bf JEL classification}\  G13

\end{abstract}

\section{Introduction}
Consider a riskless bond and a stock whose price processes
$\{B_t, t\ge0\}$ and $\{S_t, t\ge0\}$ are given by
\begin{equation}
\label{bsmarket} B_t = \exp\le(\int_0^t r(Z_s)\td s\ri),\quad  S_t
= \exp(X_t), \q X_0 = x,
\end{equation}
with $r(\cdot)\ge 0$ the instantaneous interest rate, $Z$ a finite
state Markov process  and $X=\{X_t, t\ge 0\}$ a regime-switching
phase-type L\'{e}vy process (that will be specified below in
Section \ref{sec:prel}). When $X_t$ is a Brownian motion with
drift and $r(\cdot)$ is constant, the model \eqref{bsmarket}
reduces to the classical Black-Scholes model (BS). It has been
well documented in the literature that the BS model is not
flexible enough to accurately replicate observed market call
prices simultaneously across different strikes and maturities.

To address some of the deficiencies of the BS model it was
proposed to replace the geometric Brownian motion by an
exponential L\'{e}vy process, modelling sudden stock price
movements by jumps. A substantial literature has been devoted to
the study and application of L\'{e}vy models in derivative
pricing; popular models include the infinite jump activity models,
such as the NIG \cite{NIG}, CGMY \cite{CGMY}, KoBoL \cite{KoBoL}
and hyperbolic processes \cite{EK}, and the finite activity,
jump-diffusion models -- see also Cont and Tankov \cite{CT} for an
overview. In the latter category, for instance, Kou \cite{Kou}
investigated the pricing of European and barrier options in the
case of double-exponential jumps; Asmussen et al. \cite{AAP}
considered perpetual American and Russian options under phase-type
jumps.

In a parallel line of research the BS model was extended by
allowing its parameters $\mu,\sigma$ and $r$ to be modulated by a
finite state Markov chain $Z$. The process $Z$ models (perceived)
changes in economic factors and their influence on the stock
price. See Guo \cite{GuoE,GuoQ} for background
 on this regime-switching model and further references.
 In the context of option pricing, Guo \cite{Guo,GuoE} and Guo
and Zhang \cite{GuoZ} obtained closed form solutions of European,
perpetual American put and lookback options for a two-state regime
switching Brownian motion; For the case of $N$ states, Jobert and
Rogers \cite{JR} considered the perpetual American put and
numerically solved the finite time American put problem.

In the present study we consider the model \eqref{bsmarket} which
combines both the important features of regime-switching and
jumps, motivated by the observation that L\'{e}vy models have been
successfully calibrated to options with single, short time
maturities whereas regime-switching models fit well longer dated
options. The model \eqref{bsmarket} allows, at least in principle,
for a flexible specification of the jump-distribution, since the
phase-type distributions are dense in the class of all
distributions on a half-line (see \cite[Prop. 1]{AAP}). Under this
model, we obtain explicit, analytically tractable results for the
value function of a perpetual American put and corresponding
optimal exercise strategy under this model. Guo \& Zhang
\cite{GuoZ} and Jobert \& Rogers \cite{JR} have shown that the
optimal stopping time takes the form of the first-passage problem
of $X_t$ under a level $k(Z_t)$ that depends on the current regime
$Z_t$. We will show that the optimal stopping time still takes
this form in our model and subsequently solve the corresponding
first-passage problem. The solution of the latter rests on a path
transformation and new matrix Wiener-Hopf factorization results,
which extend the classical factorization results of London et al.
\cite{LMRW}.  The results also extend Asmussen et al. \cite{AAP}
who solved the first-passage problem across a constant level in
the case of two regimes using methods different from ours.

To value a {\em finite} maturity American put under the model
\eqref{bsmarket} the solution of the perpetual American put
problem may in principle be used as building block in an
approximation procedure that we will briefly outline now. In the
setting of the BS model, Carr \cite{Carr} investigated the
approximation of a finite maturity American put price by
randomizing its maturity and showed, by numerical experiments,
fast convergence of this algorithm. The proposed maturity
randomization resulted in an iterative evaluation of a series of
related perpetual-type American options, a procedure which was
extended to the setting of jump-diffusions by Levendorskii
\cite{LQF}. The idea is then to combine our solution of the
first-passage problem with Carr's ideas to develop a pricing
algorithm of finite maturity American put options under a
regime-switching L\'{e}vy model -- we leave further exploration of
this idea for future research.

The rest of the paper is organized as follows. Section
\ref{sec:prel} is devoted to the problem formulation and the
solution of the perpetual American put problem in terms of a
first-passage problem. The solution to the first-passage problem
under a state-dependent level is developed in Sections
\ref{sec:emb}---\ref{sec:VFP}. Finally, in Section \ref{eq:AMPex},
the case of two regimes is considered in detail. Proofs that are
not given in the text are deferred to the Appendix.

\section{Problem formulation}\label{sec:prel}
\subsection{Model}
Let the bond and risky asset price processes be given as in
\eqref{bsmarket} such that $E[S_1] < \infty$, where $Z=\{Z_t;t\ge
0\}$ is a continuous time irreducible Markov process with finite
state space $E^0=\{1, \ldots, N\}$ and intensity-matrix $G$, and
$X=\{X_t, t\ge 0\}$ is a regime-switching jump-diffusion given by
\begin{equation}\label{eq:Xt}
X_t = x + \int_0^t\m(Z_s)\td s + \int_0^t\sigma(Z_s)\td W_s +
\sum_{i\in E^0}\int_0^t1_{(Z_s=i)} \td J_i(s).
\end{equation}
Here $1_B$ is the indicator of the set $B$, $x\in\R$,
$W=\{W_t;t\ge 0\}$ is a Wiener process, $J_i=\{J_i(t);t\ge 0\}$
are independent compound Poisson processes with jumps arriving at
rate $\lambda_i$, and $\m$ and $\s$ are real-valued functions on
$E^0$ with $\s(\cdot)>0$.
The stochastic processes $X$ and $Z$ are defined on some filtered
probability space $(\Omega,\mc F, \mathbb F, \P)$, where $\mathbb
F = \{\mc F_t\}_{t\ge 0}$ denotes the completed filtration
generated by $(X,Z)$.
The jump sizes of the compound Poisson processes $J_i$ are assumed
to be distributed according to {\it double phase-type
distributions}, the definition of which we will specify below. We
first briefly review the definition of a phase-type distribution.
A distribution $F$ on $(0,\infty)$ is said to be of {\it
phase-type}, if it is the distribution of the absorption time of a
finite state Markov chain with one state $\partial$ absorbing and
the remaining states transient. One writes $F\sim PH(\alpha,T)$ if
this Markov chain, restricted to the transient states, has
generator matrix $T$ and initial distribution given by the
(column) vector $\alpha$. From Markov chain theory it follows that
the density of $F$ is given by
\begin{equation}
f(x) = \alpha' \te{T x} t,\qquad x>0,
\end{equation}
where $'$ denotes transpose and $t=(-T)\mathbf 1$, with $\mathbf
1$ a column vector of ones, is the vector of exit rates from a
transient state to $\partial$. The class of phase-type
distributions is dense (in the sense of convergence in
distribution) in the class of all probability distributions on
$(0,\infty)$. Examples of phase-type distributions include
hyper-exponential and Erlang distributions. See Neuts \cite{Neuts}
and Asmussen \cite{Asm89,a,aruin} for further background on
phase-type distributions and their applications.

An extension to distributions supported on $\R$ reads as follows:
\begin{Def}
A continuous distribution $H$ on $\R$ is said to be of {\em double
phase-type with parameters $p,\a,T,\b,U$}, and one writes $H\sim
DPH(p,\a,T,\b,U)$, if its density $h$ is of the form
\begin{equation}
h(x) = p f_{\alpha, T}(x)1_{(x>0)} +
(1-p)f_{\beta,U}(-x)1_{(x<0)},
\end{equation}
where $p\in(0,1)$ and $f_{\alpha,T}$, $f_{\beta,U}$ are
$PH(\alpha,T)$ and $PH(\beta,U)$ densities respectively.
\end{Def}

For each $i\in E^0$ the jump size distribution $F_i\sim
DHP(p_i,\a^+_i,T^+_i,\a_i^-, T^-_i)$ of $J_i$ is of double
phase-type. Note that the positive and negative jumps of $J_i$
arrive at rates $\l^+_i := p_i\l_i$ and $\l_-^i := (1-p_i)\l_i$
and are distributed according to PH$(\a^+_i,T_i^+)$ and
PH$(\a^-_i,T_i^-)$ distributions, respectively.

The market with price processes $(B,S)$ as specified above is
arbitrage-free as there exists an equivalent martingale measure
$\P^*$. Furthermore, there exists a $\P^*$ that is
structure-preserving (i.e. $X$ is still of the form \eqref{eq:Xt}
but with different parameters) -- a proof of this result is given
in the Appendix. From now on we will assume that $X$ admits a
representation \eqref{eq:Xt} under a martingale measure $\P^*$,
and we will write $\P$ for $\P^*$.

\subsection{Perpetual American put}\label{eq:AMP}

In the market \eqref{bsmarket} we consider a perpetual American
put with strike $K>0$, a contract that gives its holder the right
to exercise it at any moment $t$ and receive the payment $K-S_t$.
From standard theory of pricing American style options in
\cite{bens,kara} it follows that, if $S_0=s=e^x$ and $Z_0=i$, an
arbitrage-free price for this contract is given by
\begin{equation}
\label{eq:amput} V^*(s,i) = \sup_{\t\in\mc T_{0,\i}}\E_{x,i}
\le[B_\tau^{-1}(K-\te{ X_\t})^+\ri],
\end{equation}
where $u^+ = \max\{u,0\}$, $\mc T_{0,\i}$ denotes the set of
$\mathbb F$-measurable finite stopping times, and $\E_{x,i}[\cdot]
= \E[\cdot|X_0=x,Z_0=i]$. An optimal stopping time in
\eqref{eq:amput} reads as
\begin{equation}\label{eq:Tk1k2}
T(k^*) = \inf\le\{t\ge0: X_t \leq k^*(Z_t)\ri\},
\end{equation}
for some function (or vector) $k^*:E^0\to\R$. This can be seen to
be true as follows. In view of the fact that $(S,Z)$ is a Markov
process, the general theory of optimal stopping in Shiryaev
\cite{shir} implies that an optimal stopping time $\t^*$ in
(\ref{eq:amput}) is given by $$\t^* = \inf\{t\ge0:
V^*(S_t,Z_t)\leq (K-S_t)^+\}.$$ That $\tau^*$ is of the form
\eqref{eq:Tk1k2} is a consequence of the fact that $V^*(s,i)$ is a
positive, convex and decreasing function that dominates $(K-s)^+$
with $V^*(0,i)=K$. The latter follows in turn by the definition of
$V^*$ and by observing that $s\mapsto (K-s\te{X_\t-X_0})^+$ is
convex and decreasing, and that subsequently taking the
expectation and the supremum over the stopping times $\tau\in\mc
T_{0,\i}$ preserves these properties.

The next result presents the solution to the valuation problem of
the perpetual American put in the market \eqref{bsmarket}:
\begin{Thm}\label{thm:mainap}
The value function in \eqref{eq:amput} is given by $V^*(s,i) =
V_{k^*}(s,i)$ where
\begin{equation}
V_{k}(s,i) = Kv_{0,k}(x, i) - v_{1,k}(x,i),\qquad\quad i\in E^0,
s=\te{x},
\end{equation}
with
\begin{equation*}
v_{b,k}(x,i) = \E_{x,i}\left[B^{-1}_{T(k)}\te{b X_{T(k)}}\right].
\end{equation*}
An optimal stopping time in \eqref{eq:amput} is given by
\eqref{eq:Tk1k2} where $k^*=(k^*_1, \ldots, k^*_N)$ satisfies
\begin{equation}\label{eq:smoothfit}
\lim_{x\downarrow k^*_j}V_{k^*}'(\te{x},j) = -\te{k^*_j}\quad\quad
j\in E^0.
\end{equation}
\end{Thm}

\subsection{First passage}

\begin{figure}[t]
\centering
\input{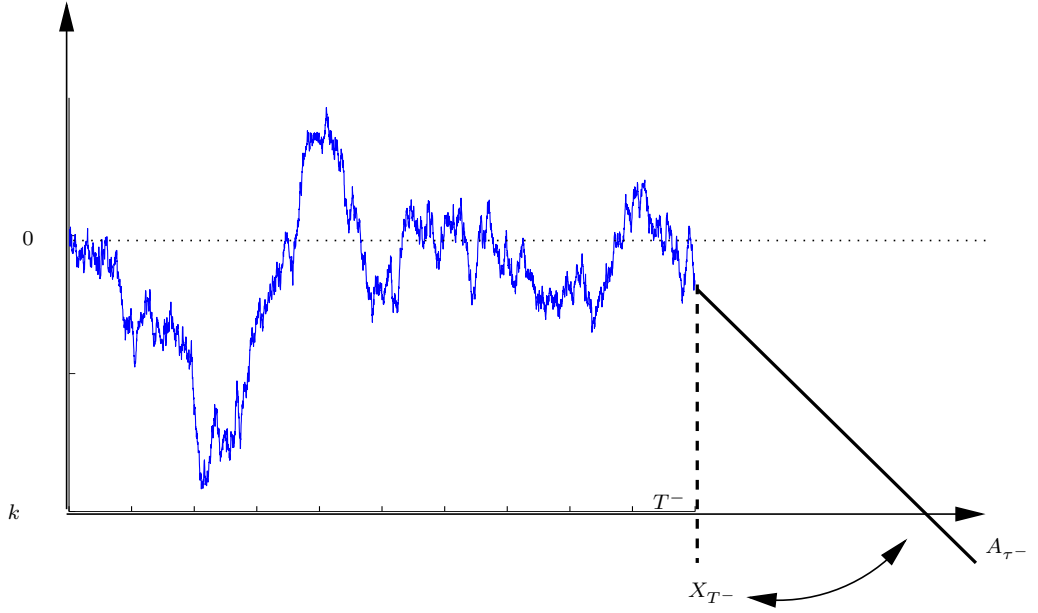}
\caption{Shown is a sample path of $X$ until the first time $T^-$
that $X$ enters $(-\i,k)$. The process $A$ has no positive jumps
and always hits a level at first-passage.} \label{fig:embed}
\end{figure}

To solve the American put problem we will consider the
first-passage problem of $X_t$ under the level $k(Z_t)$, with
$k:E^0\to\R$, which amounts to finding the function
\begin{equation}\label{eq:vipa}
v_{b,k}(x,i) = \E_{x,i}\le[\te{-R_T + bX_{T}}h_0(Z_{T})\ri],
\end{equation}
where $R_T = \int_0^T a(Z_s)\td s $, $a, h_0: E^0\to\R_+$, and
\begin{equation}
T = T(k) = \inf\le\{t\ge0: X_t \leq k\le(Z_t\ri)\ri\}.
\end{equation}
Following \cite{aruin,AAP}, the first step in the solution of
\eqref{eq:vipa} is to reformulate this problem as a first-{\em
hitting} time problem for a related {\em continuous} Markov
additive process $A$, called the {\em fluid embedding} of $X$.
Informally, a path of $A$ is constructed from a path of $X$ by
replacing the jumps of $X$ by linear stretches  -- see Figure
\ref{fig:embed}. An explicit construction is given in Section
\ref{sec:emb}.

\begin{figure}[t]
\centering
\input{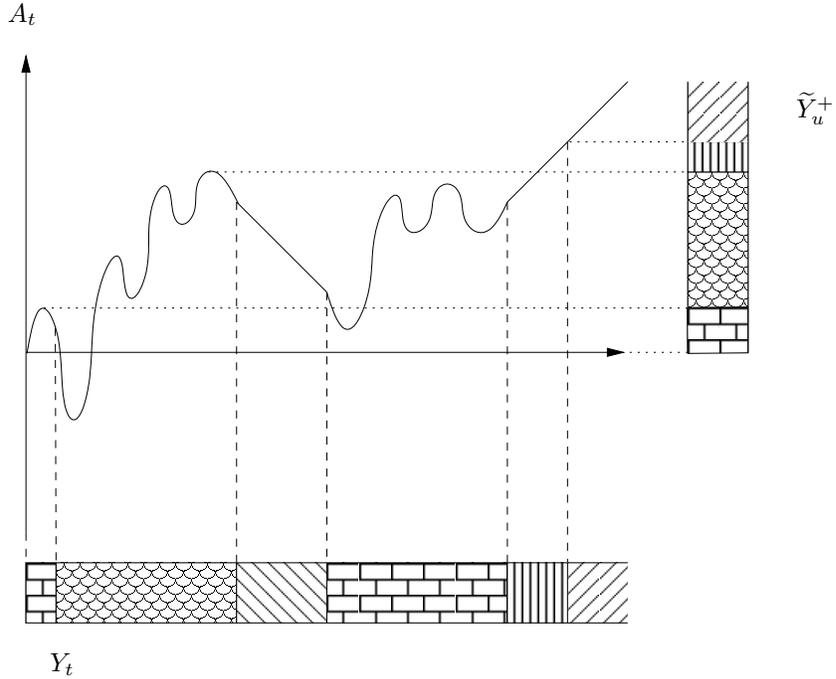}
\caption{Pictured is a stylized sample path of the process $A$.
The dashed vertical lines denote the jump times of $Y$. The
horizontal dotted lines indicate the jump times of the associated
Markov process $\WT Y^+$.} \label{fig:ladderheight}
\end{figure}

A classical approach \cite{Ke,LMRW} to solving the resulting
first-hitting time problem rests on a characterization of the laws
of corresponding up- and down-crossing ladder processes -- see
Figure \ref{fig:ladderheight}. London et al. \cite{LMRW} developed
matrix Wiener-Hopf factorization results for fluctuating additive
processes (see also Rogers \cite{rogers} for elegant martingale
proofs and Brownian perturbations). By extending the results of
\cite{LMRW,rogers} to our setting, we solve the matrix Wiener-Hopf
factorization problem for the embedding $A$, in Section
\ref{sec:MWE}.

To deal with the different ways in which first-passage in
\eqref{eq:vipa} can occur (see  Figure \ref{fig:statedep}), the
Wiener-Hopf factorization is employed in Section \ref{sec:twoex}
to calculate the distribution of the process $A$ at the first
moment of leaving a finite interval or a regime-switch, whichever
occurs first. In Section \ref{sec:VFP} the solution to
\eqref{eq:vipa} is derived by combining the foregoing results.

\section{Fluid Embedding}\label{sec:emb}

Let $Y$ be an irreducible continuous time Markov chain with finite
state space $E\cup \partial$, where $\partial$ is an absorbing
cemetery state, and denote by $A = \{A_t, t\ge 0\}$ the stochastic
process given by
\begin{equation}\label{eq:fluidem}
A_t = A_0 + \I_0^ts(Y_s)\td W_s + \I_0^t m(Y_s)\td s,
\end{equation}
where $s$ and $m$ are functions from $E\cup \partial$ to $\R$ with
$s(\partial)=m(\partial)=0$. The process $A$ is the
fluid-embedding of $X$ if the generator of $Y$ restricted to $E$
is equal to $Q_0$ where, in block notation,
\begin{equation}\label{Qa}
Q_a = \le(
\begin{array}{ccc}
 T^+  & t^+       & O    \\
 B^+  & G-D_a     & B^-  \\
 O    & t^-       & T^-  \\
\end{array}
\ri).
\end{equation}
Here $D_a$ is an $N\times N$ diagonal matrix with $(D_a)_{ii} =
\l_i + a_i$, $O$ are zero matrices of appropriate sizes and, again
in block-notation,
\begin{equation*}\label{Qa1}
B^\pm = \le(\begin{array}{ccc}
\l_1^\pm\a_1^{\pm\prime}&&\\
&\ddots&\\
&&\l_N^\pm\a_N^{\pm\prime}\\
\end{array}
\ri),\q T^\pm = \le(\begin{array}{ccc}
T_1^\pm&&\\
&\ddots&\\
&&T_N^\pm\\
\end{array}
\ri),\q t^\pm = \le(\begin{array}{ccc}
t_1^{\pm}&&\\
&\ddots&\\
&&t_N^{\pm}\\
\end{array}
\ri).
\end{equation*}
From the form of $Q_a$ it follows that $E$ can be partitioned as
$E = E^+\cup E^0\cup E^-$ where $E^0$ is the state-space inherited
from $Z$ and $E^+$ and $E^-$ are the states in which the path of
$A$ is linear with slope $+1$ or $-1$ which originate from the
positive and negative jumps of $X$, respectively. Similarly, we
will write $E_i =  E_i^+\cup\{i\}\cup E_i^-$ for the subset of $E$
corresponding to the $i$th regime of $X$ with corresponding
embedded positive and negative jumps. The functions $m(\cdot)$ and
$s(\cdot)$ are then specified as follows:
\begin{equation*}
s(j)=
\begin{cases}
\s_j & \text{if $j\in E^0$} \\ 0 & \text{otherwise}
\end{cases}
\quad\qquad m(j)=
\begin{cases}
1 & \text{if $j\in E^+$}\\
\mu_j & \text{if $j\in E^0$} \\ -1 & \text{if $j\in E^-$}
\end{cases}.
\end{equation*}

\subsection{Path transformation}
In Figure \ref{fig:embed} it is illustrated how a path of $A$ can
be transformed to obtain a path of $X$. More formally, denoting by
$$T_0(t)=\I_0^t1_{\le(Y_s\in E^0\ri)}\td s\qquad\text{and}
\qquad T_0^{-1}(u) = \inf\{t\ge0: T_0(t)>u\},
$$
the time before $t$ spent by $Y$ in $E^0$ and its right-continuous
inverse, respectively, it is not hard to verify that
$$
(A \circ T_0^{-1}, Y\circ T_0^{-1})\quad
 \text{is in law equal to} \quad
(X,Z).
$$
This implies in particular that the triplets $(T_0(\T T), A_{\T
T}, Y_{\T T})$ and $(T, X_{T}, Z_{T})$ have the same distribution,
where
$$
\T T = \T T(k) = \inf\le\{t\ge0: \text{$Y_t\in E^0$ and $A_t \leq
\T k(Y_t)$}\ri\}
$$
with $\T k: E\to \R$ given by $\T k(j) = k(i)$ for $j\in E_i$. It
is not hard to verify that state-dependent discounting (or
`killing') at rate $a(i)$ when $Y_t=i\in E^0$ can be included by
replacing $Q_0$ by the generator $Q_a$ for the vector
$a=(a(i),i\in E^0)$. Thus, it holds that
\begin{equation}\label{eq:vv}
v_{b,k}(x,i) =\E_{x,i}\le[\te{b A_{\T T}} h\le(Y_{\T T}\ri)
1_{\le(\T T < \zeta\ri)}\ri],
\end{equation}
with $Y$ now evolving according to the generator $Q_a$ and
 $\E_{i,x}[\cdot] = \E[\cdot|Y_0=i, A_0=x]$. Here,
\begin{equation}\label {eq:zeta}
\zeta = \inf\{t\ge0: Y_t \notin E\}
\end{equation}
with $\inf\emptyset = \i$, and $h: E\to \R$ is given by $h(j) =
h_0(i)$ for $j\in E_i$.

\section{Matrix Wiener-Hopf factorization}\label{sec:MWE}

The solution of the first-passage problem of the Markov process
$(A,Y)$ across a constant level is closely linked to the
{up-crossing} and {down-crossing ladder processes} $\WT Y^+$, $\WT
Y^-$  of $(A,Y)$. These processes are defined as time changes of
$Y$ that are constructed such that $Y$ is observed only when $A$
is at its maximum and at its minimum respectively, that is,
\begin{equation}
 \label{eq:jtilde} \WT Y^+_t = Y\le(\t^+_t\ri)\q
 \text{and}\q \WT Y^-_t = Y\le(\t^-_t\ri),
\end{equation}
where
$$\t^+_t = \inf\{s\ge0: A_s  > t\}\q\text{and}\q
\t^-_t = \inf\{s\ge0: A_s < t\}.$$ It is easily verified that the
ladder processes $\WT Y^+$ and $\WT Y^-$ are again Markov
processes with state spaces $E^0\cup E^+$ and $E^0\cup E^-$,
respectively. We will characterize the generators $Q^+_a$ and
$Q^-_a$ of $\WT Y^+$ and $\WT Y^-$ along with the initial
distributions $\eta^+$, defined by
\begin{eqnarray}
\label{etaijpm} \eta^+(i,j) &=& \P_{0,i}\le[\WT Y_0^+ = j,
\t_0^+<\zeta\ri]\quad \text{for $i\in E^-$, $j\in E^+\cup E^0$}
\\ \eta^-(i,j) &=& \P_{0,i}\le[\WT Y_0^- = j,
\t_0^-<\zeta\ri]\quad \text{for $i\in E^+$, $j\in E^-\cup E^0$}
\end{eqnarray}
Denote by $\mathcal Q(n)$ the set of irreducible $n\times n$
generator matrices (matrices with non-negative off-diagonal
elements and non-positive row sums) and write $\mc P(n,m)$ for the
set of $n\times m$ matrices whose rows are sub-probability
vectors. Let $V$ and $\Sigma$ denote the $|E|\times|E|$ diagonal
matrices given by $\text{diag}(m(i))$ and $\text{diag}(s(i))$,
respectively. The matrix $Q_a$ is called {\it recurrent} if its
rows sum up to zero; otherwise it is called {\it transient}. By
considering the process $A$ at the subsequent times it visits a
certain state, $r\in E$ say, and noting that this defines a random
walk, we have that in the recurrent case either $\sup_{t\ge 0} A_t
= \i$ or $\lim_{t\to\i} A_t = -\i$, $\P_{0,i}$-a.s.

Write $N, N^+$ and $N^-$ for the number of elements of $E^0, E^+$
and $E^-$, respectively and let $N_0^+=N+N^+$ and $N_0^-=N+N^-$.
We then define the Wiener-Hopf factorization of $(A,Y)$ as
follows:

\begin{Def}
Let $G^+, C^+, G^-$ and $C^-$ be elements of the sets $\mc
Q(N^+_0)$, $\mc P(N^-, N^+_0)$, $\mc Q(N^-_0)$ and $\mc P(N^+,
N^-_0)$, respectively. A quadruple $( C^+, G^+, C^-, G^-)$ is
called a {\em Wiener-Hopf factorization} of $(A,Y)$ if
\begin{equation}
\label{systemnoiseWH3} \Xi\le(-G^+, W^+\ri) =  O \q\text{and}\q
\Xi\le(G^-, W^-\ri) =  O,
\end{equation}
where, for matrices $ W$ with $|E|$ rows,
\begin{equation}\label{matrixKW}
\Xi( S, W) = \frac{1}{2}\Sigma^2 W S^2 +
 V W  S +   Q_a W
\end{equation}
and $ W^+$ and $ W^-$ are given in obvious block notation by
\begin{equation}\label{eq:wpwm}
 W^+ =
\left(
\begin{array}{cc}
 I^+ &  O \\
 O &  I_0 \\
\multicolumn{2}{c}{ C^+}
\end{array}
\right) \q\text{and}\q
 W^- =
\left(
\begin{array}{cc}
 \multicolumn{2}{c}{ C^-}\\
  I_0 &  O\\
 O &  I^-
\end{array}
\right),
\end{equation}
where $I_0$, $ I^+$ and $ I^-$ are identity matrices of sizes
$N\times N$, $N^+\times N^+$ and $N^-\times N^-$, respectively,
and $O$ denotes a zero matrix of the appropriate size.
\end{Def}

In the following result the Wiener-Hopf factorization of $(A,Y)$
is identified:

\begin{Thm}\label{thm:symemb}
(i) The quadruple $(\eta^+, Q^+,\eta^-, Q^-)$ is a Wiener-Hopf
factorization of $(A,Y)$.

(ii) {The Wiener-Hopf factorization $(\eta^+, Q^+,\eta^-, Q^-)$ is
unique if $ Q$ is transient or if $Q$ is recurrent and $A$
oscillates (that is, $\sup_t A_t=-\inf_tA_t = \i$).}

(iii) If $ Q$ is recurrent and $\lim_{t\to\i} A_t = -\i$, there
are precisely two Wiener-Hopf factorizations of $(A,Y)$ given by
$(\eta^+, Q^+,\eta^-, Q^-)$ and
$$
\le(\eta^+( I - \mathbf 1 \mu) + \mu,
 Q^+( I - \mathbf 1 \mu), \eta^-,  Q^-\ri),
$$
where $\mu$ is the left eigenvector of $ Q^+$ corresponding to its
largest eigenvalue, normalized such that $\mu \mbf 1 = 1$, where
$\mbf 1$ denotes a column vector of ones.
\end{Thm}
\medskip

\proof\  For $\ell\in\R$ let $\Phi^\pm_\ell$ be given by the
matrices
\begin{equation}\label{eq:phi+-}
\Phi^+_\ell(x) = W^+\exp\le(Q^+(\ell-x)\ri)\quad\text{and} \quad
\Phi^-_\ell(x) = W^-\exp\le(Q^-(x-\ell)\ri),
\end{equation}
where $W^+$ and $W^-$ are given by \eqref{eq:wpwm} with
$C^+=\eta^+$ and with $C^-=\eta^-$. The proof rests on the
martingale property of
\begin{equation}\label{eq:M+-}
M^+_t = f_+\le(Y_{t\wedge\t_\ell^+},A_{t\wedge\t^+_\ell} \ri)\quad
\text{and}\quad M^-_t =
f_-\le(Y_{t\wedge\t^-_\ell},A_{t\wedge\t^-_\ell} \ri)
\end{equation}
with
\begin{equation}\label{eq:f+-}
f_+(i,x) = e_i'\Phi^+_\ell(x)h_+\quad\text{and}\quad f_-(i,x) =
e_i'\Phi^-_\ell(x)h_-,
\end{equation}
where $h_+$ and $h_-$ are $N_0^+-$ and $N_0^--$ column vectors,
respectively. Here, and in the sequel, $e_i$ denotes a (column)
vector of appropriate size with element $e_i(j) = 1$ if $j=i$ and
zero otherwise. To verify that $M^+$ is a martingale, observe
first that, since $\WT Y^+$ is a Markov process with generator
$Q^+=Q^+_a$ and initial distribution $\eta^+$ defined by
(\ref{etaijpm}), Markov chain theory implies that for $x\leq \ell$
\begin{equation}\label{eq:ffexp}
\E_{x,i}\le[h\le(\WT Y^+_\ell\ri)1_{\le(\t^+_\ell < \zeta\ri)}\ri]
= e_i'\Phi^+_\ell(x) h.
\end{equation}
The martingale property of $M^+$ then follows from \eqref{eq:f+-}
-- \eqref{eq:ffexp} as a consequence of the Markov property of
$(A,Y)$. An application of It\^o's lemma shows that $f =
(f_+(i,u),i\in E)$ satisfies for $u\leq \ell$
\begin{equation}
\label{ITOPHASE2} \mbox{$\frac{1}{2}$}s(i)^2 f''(i,u) + m (i)
f'(i,u) + \sum_{j}q_{ij}(f(j,u)-f(i,u)) =  0,
\end{equation}
where $ f'$ and $ f''$  denote the first and second derivatives of
$f$ with respect to $u$.
By substituting the expressions (\ref{eq:phi+-}) -- \eqref{eq:f+-}
into equation (\ref{ITOPHASE2}) we find, since $ h$ was arbitrary,
that $ Q^+$ and $\eta^+$ satisfy the first set of equations of the
system (\ref{systemnoiseWH3}). The proof for $ Q^-$ and $\eta^-$
is analogous and omitted. The proofs of Theorem \ref{thm:symemb}
(ii), (iii) are deferred to the Appendix.\exit
\bigskip

\noindent{\bf Example (Gerber-Shiu penalty function)} Let $X_t$ in
\eqref{eq:Xt} with $x>0$ be the surplus of an insurance company.
Note that, for $\s=0$, $N=1$, and in the absence of positive
jumps, \eqref{eq:Xt} reduces to the classical Cram\'{e}r-Lundberg
model. Denote by $\rho=\inf\{t\ge0: X_t<0\}$ the ruin time of $X$.
Quantities of interest in this setting include the ruin
probability $P_{x,i}(\rho<\infty)$ and the Gerber-Shiu \cite{GS}
expected discounted penalty function which quantifies the severity
of ruin by measuring the shortfall $X_\rho$ of $X$ at the ruin
time $\rho$. Under the model \eqref{eq:Xt} both these quantities
can be expressed in terms of the functions $\Phi^\pm_\ell$ defined
in \eqref{eq:f+-}. For instance, the probability of ruin in regime
$\ell\in E^0$ is given by
$$
\P_{x,i}\le(\rho < \i, Z_\rho = \ell\ri) = e_i'\Phi^-_0(x) f_\ell,
$$
where  $f_\ell(j) = 1$ if $j\in E_\ell$ and zero else and
$\Phi^-_0$ is given by \eqref{eq:phi+-} with $Q^-=Q^-_0$. More
generally, for any non-negative function $\pi$ on $[0,\i)\times
E^0$ the Gerber-Shiu
 expected discounted penalty function reads as
$$
\E_{x,i}\le[\te{-R_\rho}\pi(X_\rho, Z_\rho)\ri] =
e_i'\Phi^-_0(x)\bar g,
$$
where $R_\rho = \int_0^\rho a(Z_s)\td s$, $\Phi^-_0$ is given by
\eqref{eq:phi+-} with $Q^-=Q^-_a$, and $\bar g$ is the
$N^-$-vector with elements
$$
\bar g(\ell) = \begin{cases}
\pi(0,\ell) & \text{for $\ell\in E^0$}\\
\displaystyle\int_0^\i \pi(-s,m) e_\ell'\;\te{sT_m^-}t_m^-\td s
&\text{for $\ell\in E^-_m$}\\
\end{cases}.
$$

\section{First exit from a finite interval}\label{sec:twoex}
The two-sided exit problem of $A$ from the interval $[k,\ell]$ for
$-\infty<k<\ell<+\infty$ is to find the distribution of the
position of $(A_{\t},Y_{\t})$ at the first-exit time
$$\t =
\t_{k,\ell}=\inf\{t\ge0: A_t\notin[k,\ell]\}.$$
By considering appropriate linear combinations of the martingales
$M^+$ and $M^-$ defined in \eqref{eq:M+-} we will now show that
the two-sided exit problem can be solved explicitly in terms of
$(\eta^+, Q^+,\eta^-, Q^-)$. To this end, introduce
$$
 Z^+ =  \left(
\begin{array}{cc}
 O &  I_0\\
\multicolumn{2}{c}{ \eta^+}\\
\end{array}
\right)\te{ Q^+(\ell-k)}, \quad  Z^- =  \left(
\begin{array}{cc}
\multicolumn{2}{c}{ \eta^-}\\
 I_0 &  O\\
\end{array}
\right) \te{ Q^-(\ell-k)},
$$
and define the matrices
\begin{eqnarray}\label{eq:twosidedplus}
\Psi^+(x) &=& \le( W^+\te{ Q^+(\ell-x)} - W^-\te{ Q^-
(x-k)} Z^+\ri) \le( I -  Z^- Z^+\ri)^{-1},\\
\label{eq:twosidedmin} \Psi^-(x) &=& \le( W^-\te{ Q^-(x-k)} -
W^+\te{ Q^+
(\ell-x)} Z^-\ri) \le( I -  Z^+ Z^-\ri)^{-1},\\
\label{eq:theta} \Psi^\circ(s,x) &=& \le(\te{sx} I - \te{s
\ell}\Psi^+(x)J^+ - \te{s k} \Psi^-(x) J^-\ri)[-K(s)]^{-1},
\end{eqnarray}
where $I$ is an identity matrix, $W^\pm$ are given in
(\ref{eq:wpwm}) with $ C^+=\eta^+$ and $ C^-=\eta^-$, $J^\pm$ is
the transpose of \eqref{eq:wpwm} with $C^\pm$ replaced by zero
matrices, and
\begin{equation}\label{eq:K[s]}
K(s) = \frac{1}{2}\Sigma^2s^2 + Vs + Q_a.
\end{equation}
 We will write $\Psi^{+}_{k,\ell}/\Psi^{-}_{k,\ell}/\Psi^{\circ}_{k,\ell}$
if we wish to clarify their dependence on $k$ and $\ell$.

The complete solution of the two-sided exit problem reads as
follows:
\begin{Prop}\label{prop:twoexit}
Let $h^+, h^-$ and $h^\dagger$ be functions that map $E^0\cup
E^+$, $E^0\cup E^-$ and $E$ to $\R$. If
\begin{equation} s(i)^2b^2 + m(i)b < -q_{ii}\qquad \text{for all $i\in
E$},
\end{equation}
then it holds for $x\in[k,\ell]$ and $i\in E$ that
\begin{eqnarray}\label{eq:psi-up}
\E_{x,i}\le[h^+\le(Y_\t\ri)1_{\le(A_\t = \ell, \t<\zeta\ri)}\ri]
&=& e_i'\Psi^+(x) h^+,\\ \label{eq:psi-down}
\E_{x,i}\le[h^-\le(Y_{\t}\ri)1_{\le(A_\t = k, \t<\zeta\ri)}\ri]
&=& e_i'\Psi^-(x) h^-,\\ \label{eq:psi-dagger}
\E_{x,i}\le[\te{bA_{\zeta-}}h^\dagger\le(Y_{\zeta-}\ri)1_{\le(\zeta<\tau\ri)}\ri]
&=& e_i'\Psi^\circ(b,x)\Delta_{h^\dagger}q_a,
\end{eqnarray}
where $q_a = (-Q_a)\mbf 1$, $\zeta$ is defined by ~\eqref
{eq:zeta} and $\Delta_{h^\dagger}$ is the diagonal matrix with
elements $h^\dagger(j)$.
\end{Prop}
\proof\ Define $g_+$ and $g_-$ by the right-hand sides of
\eqref{eq:psi-up} and \eqref{eq:psi-down} respectively. It is
straightforward to verify from \eqref{eq:twosidedplus} --
\eqref{eq:twosidedmin} that it holds that
\begin{eqnarray*}
g_+(i,x) &=& \begin{cases}h^+(i) & \text{if $x=\ell, i\in E^+\cup
E^0$}\\ 0 & \text{if $x=k, i\in E^-\cup E^0$}
\end{cases},\\
g_-(i,x) &=& \begin{cases}0 & \text{if $x=\ell, i\in E^+\cup
E^0$}\\ h^-(i) & \text{if $x=k, i\in E^-\cup E^0$}
\end{cases}.
\end{eqnarray*}
In view of these boundary conditions and the fact that any linear
combination of $M^+$ and $M^-$, defined in \eqref{eq:M+-}, is a
bounded martingale, Doob's optional stopping theorem gives that
\begin{eqnarray*}
g_+(i,x) &=& \E_{x,i}\le[g_+(Y_{\t}, A_\t)1_{(\tau<\zeta)}\ri] \\
&=& \E_{x,i}\le[h^+(Y_{\t})1_{\le(A_{\t} = \ell,
\tau<\zeta\ri)}\ri],
\end{eqnarray*}
where $\t=\t_{k,\ell}$. Similarly, it follows that
$$g_-(i,x) = \E_{x,i}\le[h^-(Y_{\t})1_{\le(A_{\t} = k, \tau<\zeta\ri)}\ri].$$
To prove the third identity, consider the map $h^*:E\to\R$ given
by $$h^*(i) = \E_{0,i}\le[\te{s
A_{\zeta-}}h^\dagger\le(Y_{\zeta-}\ri)\ri].$$ By conditioning on
the first jump epoch $\xi$ of $Y$ it is straightforward to verify
that
\begin{eqnarray*}
h^*(i) 
&=& - \frac{h^\dagger(i)q_{i\partial} + \sum_{j\neq
i}q_{ij}h^*(j)}{s(i)^2s^2/2 + m(i) s + q_{ii}},
\end{eqnarray*}
where $q_{ij} [q_{i\partial}]$ denotes the intensity of a
transition $i\to j$ $[i\to\partial]$. After reordering and writing
the above expression in matrix form, it follows that $$ K(s) h^* =
\Delta_{h^\dagger}Q_a\mbf 1,$$ so that, for $s$ satisfying
$s(i)^2s^2 + m(i)s < -q_{ii}$, $i\in E$,
\begin{equation}\label{eq:hstar}
h^* = [-K(s)]^{-1}\Delta_{h^\dagger}(-Q_a)\mbf 1.
\end{equation}
In view of the strong Markov property and \eqref{eq:psi-up} --
\eqref{eq:psi-down}, it follows that
\begin{eqnarray}
\nn \E_{x,i}\le[\te{s A_{\zeta-}}1_{(\zeta < \t)}\ri] &=&
\E_{x,i}\le[\te{s A_{\zeta-}}\ri] -
\E_{x,i}\le[\te{s A_{\zeta-}}1_{(\zeta > \t)}\ri]\\
\nn &=&  \te{sx}h^*(i) - \E_{x,i}\le[\te{s A_{\t}}1_{(\t < \zeta)}
h^*(Y_{\t})\ri]\\
&=& e_i'\le(\te{sx}h^* - \te{sk}\Psi^-(x)J^- h^* -
\te{s\ell}\Psi^+(x)J^+h^*\ri). \label{eq:toos}
\end{eqnarray}
Inserting \eqref{eq:hstar} into \eqref{eq:toos} finishes the proof
of \eqref{eq:psi-dagger}.\exit

\section{First-passage under state-dependent levels}\label{sec:VFP}

\begin{figure}[t]
\centering
\input{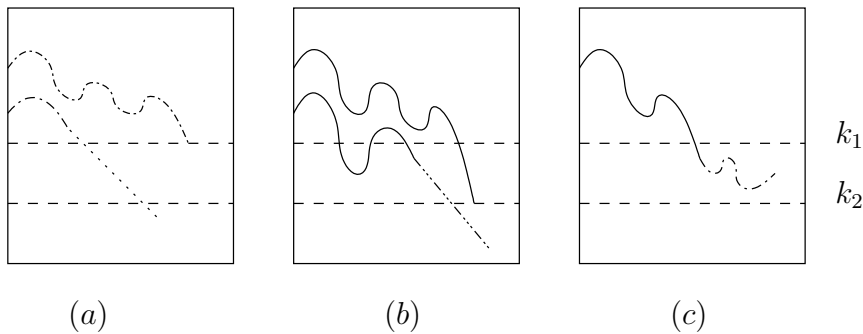}
\caption{First-passage under the state-dependent  level
$(k_2,k_1)$ takes place while $Y\in E^0$ and can take place in two
ways: $A$ hits the level $k_i$ while $Y=i$, $i=1,2$ (illustrated
in (a) and (b)) or by a jump of $Y$ (illustrated in (a), (b) and
(c)). The different line styles of the paths of $A$ correspond to
the different states of $Y$. } \label{fig:statedep}
\end{figure}
By combining the ingredients from the previous sections the
first-passage function $v_{k,b}(x,i)$ of $A_t$ under the level
$k(Y_t)$
\begin{equation}\label{eq:vvv}
v_{b,k}(x,i) = \E_{x,i}\le[\te{b A_{\T T(k)}} h\le(Y_{\T
T(k)}\ri)1_{\le(\T T(k) < \zeta\ri)}\ri],
\end{equation}
with $h: E\to\R_+$ and $$\T T(k)=\inf\le\{t\ge0: \text{$Y_t\in
E^0$ { and}  $A_t\leq \T k(Y_t)$}\ri\}$$ can be explicitly
expressed in terms of the matrix Wiener-Hopf factorization found
in Theorem \ref{thm:symemb}. For simplicity we will assume that
the levels are ordered as $k_1 > k_2 > \ldots > k_N$ (the general
case of possibly equal levels follows by a similar reasoning). As
the first-passage over $k$ can only occur when $Y$ is in $E^0$, it
follows that $Y$ can cross the boundary $k$ {\em before} $A$ exits
the interval $[k_j, k_{j-1}]$ in two ways: either $Y$ jumps into a
state $\{1, \ldots, j-1\}$ or $A$ hits the level $k_j$ while $Y$
is in state $j$ -- see Figure \ref{fig:statedep}. We are thus led
to considering the processes
$$
\le.Y^{(j)} =  Y\ri|_{\WT E_j}, \qquad\text{where $\WT E_j:=
E\backslash \cup_{i=1}^{j-1} E_i$}
$$
(with $Y^{(1)}=Y$). Clearly, the $Y^{(j)}$ are themselves Markov
processes with generators $Q_{(j)}$ given by the corresponding
restrictions of $Q_a$; in block notation the resulting partitions
read as
$$
Q_a = \le(
\begin{array}{cc}
\multicolumn{2}{c}{R^{(j)}}\\
q^{(j)} & Q_{(j)}\\
\end{array}
\ri),\quad\quad\quad j=2, \ldots, N,
$$
for some matrix $R^{(j)}$, where $q^{(j)}$ is the matrix of exit
rates from the sub-space $\WT E_j$. By the strong Markov property,
Proposition \ref{prop:twoexit} and Theorem \ref{thm:symemb}, the
value of $v_{b,k}(x,i)$ can be expressed in terms of the unknowns
$v_{b,k}(k_j,i)$. For these unknowns a system of equations can be
derived by invoking smoothness and continuity properties of
$v_{b,k}$ above the barrier $k$.

More specifically, denote, for some constants $C_{j}(\ell),
D_j(\ell)$, the vectors $h^-_{j}$ and $h^+_{j}: \WT E_j\to\R$ by
\begin{eqnarray}\label{eq:h+}
h^+_{j}(\ell) &=& \begin{cases}
C_{j-1}(\ell)   &\text{if $\ell\in E^0\backslash\{1, \ldots, j-1\}$}\\
D_{j-1}(\ell)  &\text{else}
\end{cases},\quad\quad (j=2, \ldots, N),\\
\label{eq:h-}
h^-_{j}(\ell) &=& \begin{cases}\te{b k_j}h(j) & \text{if $\ell=j$}\\
C_j(\ell) &\text{if $\ell \in E^0\backslash\{1,\ldots, j\}$}\\
D_j(\ell) &\text{else}
\end{cases},\quad\quad (j=1,\ldots, N),
\end{eqnarray}
respectively, and set $$h^\dagger_{j} = (h(\ell), \ell\in\WT
E_j).$$ We shall write $\Psi^+_{j}$, $\Psi^-_{j}$ and
$\Psi^\circ_{j}$ as shorthand for $\Psi^+_{(k_j,k_{j-1})}$,
$\Psi^-_{(k_j,k_{j-1})}$ and $\Psi^\circ_{(k_j,k_{j-1})}$,
respectively, and denote by $f(x+)$ and $f(x-)$ the right- and
left-limits of the function $f$ at $x$. Then the following
characterization of $v$ holds true:

\begin{Thm}\label{thm:fipakz}Assume
that $s(i)^2b^2/2 + m(i) b < -q_{ii}$ for all $i\in E$. The
function $v_{b,k}$ is given by
\begin{equation}
v_{b,k}(x,i) =
\begin{cases}
e_i' \Phi^-(x) h^-_{1}, &\text{if $x>k_1$,}\\
\\
e_i'\le[\Psi^+_{j}(x)h_{j}^+ + \Psi^-_{j}(x)h_{j}^- +
\Psi^\circ_{j}(b,x) \Delta_{h_{j}^\dagger} q^{(j)}\ri], &
\text{if $j=2,\ldots, N,$} \\
& \text{$k_j < x \leq k_{j-1}$,}
\end{cases}
\label{eq:vvvv}
\end{equation}
where $\Phi^-=\Phi^-_{k_1}$ is given by \eqref{eq:f+-} and
$h^+_{j}, h^-_{j}$ are specified by \eqref{eq:h+} -- \eqref{eq:h-}
with $C_{j}(\ell)$ and $D_{j}(\ell)$ satisfying the following
system of equations:
\begin{eqnarray}
v_{b,k}'( k_j + , \ell ) &=& v_{b,k}'( k_j -, \ell), \quad\quad
\ell\in E^0\backslash\{1, \ldots, j\},
\label{eq:vC} \\
v_{b,k}( k_j + , \ell ) &=& v_{b,k}( k_j - , \ell), \quad\quad
\ell\in\WT E_j\le\backslash\le[E^0\cup E_j^-\ri]\ri.,
\label{eq:VD}\\
\le(D_{j}(\ell), \ell\in E_j^-\ri) &=& \te{bk_j}\le(bI -
T_j^-\ri)^{-1}t^-_jh(j), \label{eq:DD}
\end{eqnarray}
where $j=1, \ldots, N-1,$ and $'$ denotes differentiation with
respect to $x$.
\end{Thm}

\section{Example: two regimes}\label{eq:AMPex}

To illustrate the results derived in previous sections we consider
next the model \eqref{eq:Xt} in the case of two regimes.
Suppose that $Z$ is a Markov chain with state space $E^0 =
\{1,2\}$ and transition matrix $$ G = \le(\begin{array}{cc}
          -q_1 & q_1 \\
           q_2 & -q_2
 \end{array}\ri)$$
and that $X$ evolves as a Brownian motion with drift $\mu_1 t +
\sigma_1 W_t$ when $Z$ is in state $1$ and as the jump-diffusion
$\mu_2 t + \sigma_2 W_t - J_t$ when $Z$ is in state $2$, with $J$
a compound Poisson process with intensity rate $\l$ and
exponential jumps with mean $1/\alpha$. Then the embedding of
$(X,Z)$ has state space $E = \{1,2,2^\star\}$, say, with
corresponding transition matrix

\begin {equation} \label {killgen}
Q_r =\begin{pmatrix}
  - q_1 - r_1 & q_1 & 0 \\
  q_2 & - q_2 - r_2 - \lambda & \lambda \\
  0 & \alpha & - \alpha
\end{pmatrix}.
\end {equation}
We will consider the stopping time $T(k^*_1, k^*_2)$ for the three
different configurations of the optimal levels: $k^*_1 < k^*_2$,
$k^*_1=k^*_2$ and $k^*_1>k^*_2$. For $x>\max\{k_1^*,k_2^*\}$, the
value function of the put is determined by the generator matrix
$Q^-_r$ of the corresponding down-crossing ladder process, which
we determine by invoking the matrix Wiener-Hopf factorization
results from Section \ref{sec:MWE}. Noting that $E^+=\emptyset$
and $E^-=\{2^\star\}$, it follows that $Q^-=Q^-_r$ satisfies

\begin{equation}\label{eq:matqq}
\frac{1}{2}\Sigma^2 (Q^-)^2 + VQ^- + Q_r = O,
\end{equation}
where
$$
\Sigma =
\begin{pmatrix}
\sigma_1 & 0 & 0\\
0 & \sigma_2 & 0\\
0 & 0 & 0
\end{pmatrix},
\qquad\qquad V =
\begin{pmatrix}
\mu_1 & 0 & 0\\
0  & \mu_2 & 0\\
0  & 0 & -1
\end{pmatrix}.
$$
Denoting by $\b[\th]$ a right-eigenvector of $Q_r^-$ corresponding
to eigenvalue $\th$ it follows by right-multiplying
\eqref{eq:matqq} with $\beta[\th]$ that the matrix $$K(\th) =
\frac{1}{2}\Sigma^2\th^2 + V\th + Q_r$$ is singular and
$K(\th)\b[\th] = 0$. It is a matter of algebra to verify that
$\th$ satisfies $g(\th) = 0$ where

\begin {equation} \label{pol}
g(\theta)=F_1(\theta)\le((\alpha + \theta) F_2(\theta) -
\lambda\theta\ri)-q_1q_2(\alpha + \theta),
\end {equation}
with
\begin {equation} \label {meq2}
F_j(\theta)=\frac 12 \sigma_j^2\theta ^2 + \mu_j \theta -q_j-r_j,
\quad j=1,2.
\end {equation}
The following result lists the properties of the roots of $g(\th)
= 0$:

\begin{Lemma} \label {le11}
Suppose that $r_1,r_2>0$. Then $g(\theta)$ has five different real
roots which satisfy the distribution characteristics
$\theta_1<\theta_2 <\theta_3 < 0 <\theta_4<\theta_5$.
As a consequence, $Q^- = Q^-_r$ has three distinct eigenvalues
$\theta_1$, $\theta_2$ and $\theta_3$.
\end{Lemma}
Since the eigenvectors $\beta[\th_i]$ corresponding to the
different eigenvalues $\th_i$, $i=1,2,3$, are linearly
independent, the matrix $Q^-$ explicitly reads as
\begin {equation}\label{eq:q-}
 {Q}^- =
\begin{pmatrix}
 \theta_1\beta [\theta_1] & \theta_2 \beta [\theta_2]
& \theta_3 \beta[\theta_3]
\end{pmatrix}
{\begin{pmatrix}
  \beta [\theta_1] & \beta [\theta_2]
& \beta [\theta_3]
\end{pmatrix} }^{-1}.
\end {equation}

{\bf 1. Case $k=k^*_1=k^*_2$}.\q For $x>k$, the value function of
the American put reads as
$$
W(\te{x},i) = e_i' \te{Q^-(x-k)} H(k)  \quad\text{with}\q H(k) =
\begin{pmatrix}
K - \te{k}\\
K - \te{k}\\
K - \te{k}\frac{\alpha}{\alpha + 1}
\end{pmatrix},
$$
where $Q^-$ is given in \eqref{eq:q-} and $k$ solves
$$
-\te{k} = e_i' Q^- H(k)\quad\quad i=1,2.
$$

{\bf 2. Case $k_2^*>k_1^*$}.\q To deal with the case that $k^*_1 <
x< k^*_2$ and $Z=1$, we note that, if the process $Z$ is
restricted to state 1, $X$ is equal to a Brownian motion with
drift, $\m_1 t + \s_1 W_t$, killed at rate $q_1 + r_1$. The
generator matrices of this restriction of $Z$ and the
corresponding ladder processes, denoted by $\bar Q,-\bar Q^+, \bar
Q^-$, reduce in this case to scalars, given by
$$\bar Q=
-q_1-r_1$$
and the positive and negative root of the equation
$$
\frac{1}{2}\s_1^2 x^2  + \m_1 x - q_1 - r_1 = 0.
$$
The associated two-sided exit probabilities from the interval
$[k_1^*, k_2^*]$ are
\begin{eqnarray*}
\Psi^-_1(x) &=& \frac{\te{\bar Q^-(x-k)} -
\te{\bar Q^+(\ell-x)}\te{\bar Q^-(\ell-k)}}%
{1 - \te{\bar Q^+(\ell-k)}\te{\bar Q^-(\ell-k)}},\\
\Psi^+_1(x) &=& \frac{\te{\bar Q^+(\ell-x)} - \te{\bar Q^-(x-k)}
\te{\bar Q^+(\ell-k)}}%
{1 - \te{\bar Q^-(\ell-k)}\te{\bar Q^+(\ell-k)}},
\end{eqnarray*}
with $k=k_{1}^*$ and $\ell = k_2^*$. Putting everything together
shows that the value function of the American put in this case is
given by
\begin{equation*}
W(\te{x},i) = \begin{cases}
e_i' \te{Q^-(x-k^*_2)}\bar  H(k^*_2), & x \ge k_2^*,\ i=1,2,\\
\\
H^+(x) + \Psi^+_{1}(x) [C - H^+(k^*_2)] + \Psi^-_{1}(x)
H^-(k^*_1), & k_1^* < x \leq k_2^*,\ i=1,
\end{cases}
\end{equation*}
where $$ \bar  H(k) =
\begin{pmatrix}
C\\
K-\te{k}\\
K - \te{k}\frac{\alpha}{\alpha+1}
\end{pmatrix},\q
H^-(k) = \te{k} \frac{r_1}{q_1}, \q H^+(x) = K  - \te{x} \frac{q_1
+ r_1}{q_1},$$ using equation \eqref{eq:coefc}. Here $C$ is
determined by
$$
W'(\te{k_2}-,1)= e_1'Q^-\bar  H(k_2)$$ and the levels $k_1^*$ and
$k_2^*$ satisfy the smooth fit equations
\begin{eqnarray*}
\Psi^{-\prime}_1(k_1)H^-(k_1)+\Psi^{+\prime}_1(k_1)[C-H^+(k_2)]
&=& \frac{r_1}{q_1} \te{k_1}, \\  e_2'Q^-\bar H(k_2) &=& -
\te{k_2},
\end{eqnarray*}
where prime in the first equation denotes differentiation with
respect to $x$.

{\bf 3. Case $ k_1^* > k_2^* $}\q For $k^*_2 < x < k^*_1$ and
$Z=2$, we are led to consider the Markov process $Y^{(2)}$ with
state space $\{2,2^\star\}$ and generator matrix
$$
Q_{(2)} = \le(
\begin{array}{cc}
-q_2 - r_2 -\l & \l\\
\a & -\a
\end{array}
\ri).
$$
In this case it can be checked from ~\eqref {systemnoiseWH3},
~\eqref {matrixKW}, and ~\eqref {eq:wpwm} with $Q_a$ replaced by
$Q_{(2)}$, that $-Q_{(2)}^+$ is a scalar given by the positive
root of
$$
\frac{\s^2_2}{2}x^2 + \m_2 x + \lambda\frac{\a}{x+\a} = q_2+r_2+\l
$$ and that $$\eta^+ = \a/[-Q_{(2)}^+ + \a].$$
We can calculate $Q_{(2)}^-$ in a similar way as we calculated
$Q^-$ above. Writing $\T Q^+ = Q_{(2)}^+$ and $\T Q^- = Q_{(2)}^-$
the corresponding matrices of two-sided exit probabilities from
$[k_2^*,k_1^*]$ read as
\begin{eqnarray*}
\Psi^+_2(x) &=& \frac{1}{c}\le[\begin{pmatrix} 1\\ \eta^+
\end{pmatrix}
\te{\T Q^+(\ell - x)} - \te{\T Q^-(x-k)} \begin{pmatrix} 1\\
\eta^+
\end{pmatrix}
\te{\T Q^+(\ell - k)}\ri],\\
\Psi^-_2(x) &=& \le[\te{\T Q^-(x-k)} - \te{\T Q^+(\ell - x)} M
\te{\T Q^-(\ell - k)}\ri] \le[I - \te{\T Q^+(\ell - k)} M \te{\T
Q^-(\ell - k)}\ri]^{-1},
\end{eqnarray*}
with $k=k_2^*$ and $\ell=k_1^*$, where
\begin{eqnarray*}
c &=& 1 - (\ 1 \quad 0\ )\te{\T Q^-(\ell - k)}\begin{pmatrix} 1\\
\eta^+
\end{pmatrix}\te{\T Q^+(\ell - k)},\\
 M &=& \begin{pmatrix} 1 & 0\\ \eta^+ & 0
\end{pmatrix}.
\end{eqnarray*}
The value function is
\begin{equation*}
W(\te{x},i) = \begin{cases}
e_i' \te{Q^-(x-k^*_1)}\WT H(k^*_1), & x \ge k_1^*,\ i=1,2,\\
\\
\WT H^+(x) + f\le[\Psi^+_{2}(x) [C - \WT H^+(k^*_1)] +
\Psi^-_{2}(x) \WT H^-(k^*_2)\ri], & k_2^* < x \leq k_1^*,\ i=2,
\end{cases}
\end{equation*}
where $f$ is the row vector $f=(\ 1 \quad 0\ )$ and
$$
\WT H(k) =
\begin{pmatrix}
K-\te{k}\\
C\\
D
\end{pmatrix},
\q \WT H^+(x) = K - \te{x} \frac{q_2 + r_2}{q_2}, \q \WT H^-(k) =
\te{k}\frac{r_2}{q_2}\; \begin{pmatrix}
1\\
\frac{\alpha}{\alpha+1}
\end{pmatrix},
$$
where we used again equation \eqref{eq:coefc}. Here $C$ and $D$
are determined by the two linear equations
\begin{eqnarray*}
W(\te{k_1} -, 2^\star) &=& W(\te{k_1} +, 2^\star), \\
W'(\te{k_1} -,2) &=& W'(\te{k_1} +,2),
\end{eqnarray*}
and the levels $k^*_1$ and $k^*_2$ satisfy the two equations
\begin{eqnarray*}
f\le[\Psi^{-\prime}_2(k_2)\WT H^-(k_2)+\Psi^{+\prime}_2(k_2)
\{C-\WT H^+(k_1)\}\ri] &=& \te{k_2}\frac{r_2}{q_2},\\
e_1'Q^-\WT H(k_1) &=& -\te{k_1}.
\end{eqnarray*}

\subsection*{Acknowledgements} We would like to thank Florin Avram
for inspiring conversations. We also thank an anonymous referee
and an associate editor for their numerous useful suggestions and
comments that led to considerable improvements of the presentation
of the paper.

\bigskip
\renewcommand{\theequation}{A.\arabic{equation}}
\renewcommand{\thesection}{A}
\renewcommand{\thesubsection}{A.\arabic{subsection}}
\setcounter{equation}{0}  
\setcounter{section}{0}%
\setcounter{subsection}{0}%
\centerline{\bf\large Appendix}  
\section{Proofs}
\subsection{Cram\'{e}r Martingale Measure}
{ In this section and the next we present a construction of an
equivalent martingale measure for the process $(X,Z)$ and show how
the parameters change under this change of measure. For a
background of Markov additive processes we refer to Asmussen
\cite{aruin}.

An important role in the construction of the change of measure is
played by the process $X_{a} = \{X_{a}(t);t\ge0\}$ defined by
$$X_{a}(t) = \I_0^ta(Z_s)\td X_s$$ for some function $a$ to
be specified below. It is straightforward to verify that the
process $X_{a}$ is still of the form \eqref{eq:Xt}, but with
changed parameters; its characteristic matrix is $K_{a}[s]= G +
\D_{a}[s]$, where $\D_{a}[s]$ is the diagonal matrix with elements
$\k_i(a_i s)$. Write $h$ and $\l$ for the Perron-Frobenius
right-eigenvector and eigenvalue of $ K_{a}[1]$, respectively, and
define the candidate change of measure $L=\{L_t; t\ge 0\}$ by
\eqn{\label{eq:martL} L(t) = \te{X_{\Bf a}(t) - \l
t}h(Z(t))/h(Z(0)), } with $h(i)= h_i$ the $i$th coordinate of $
h$. If $\E[S_1]<\i$, then $\kappa_i(1)<\i$ for $i\in E^0$, and a
solution $a_i$, $i\in E^0$, exists of the equations
\begin{equation}
\k_i(a_i+1) = r(i) + \k_i(a_i),
\end{equation}
with $\kappa_i(a_i)<\i$, where $\k_i(s)=\log E[\te{s X_1^i}]$ is
the Laplace exponent of $X^i_t=\mu_i t + \s_i W_t + J_i(t)$. It is
shown in the following result that the measure $\P^*$ with
Radon-Nikodym derivative $$ \le.\frac{\td\P^*}{\td\P}\ri|_{\mc
F_t} = L(t) $$ is indeed an equivalent martingale measure:

\begin{Prop}\label{EMM}
Suppose that $\E[S_1]$ is finite.

(i) The process $L = \{L_t,\mc F_t; t\ge0\}$ is a positive mean
one martingale and $\P^*$ is a probability measure;

(ii) Under $\P^*$, $\exp\le(X_t- \int_0^tr(Z_s)\td s\ri) =
B_t^{-1}S_t$ is a  martingale.
\end{Prop}

In view of  Proposition \ref{EMM}, the market \eqref{bsmarket}
with price processes as specified in \eqref{eq:Xt} is
arbitrage-free if $\E[S_1]<\i$. It is shown below in Proposition
\ref{prop:com} that, under $\P^*$, $X$ is still of the form
(\ref{eq:Xt}) but with changed parameters. The process
$B_t^{-1}S_t$ is martingale if the following restriction holds for
the parameters of $X$ (see also \cite[Sec. 2]{AAP}):
\begin{equation}\label{eq:coefc}
\frac{\s^2(i)}{2} + \m(i) + \l_i (\WH F_i(1) - 1) =
r(i),\quad\quad i\in E^0,
\end{equation}
where
$$\WH F_i(s)=
p_i\a_i'(-sI - T_i)^{-1}t_i + q_i\beta_i'(sI - U_i)^{-1}u_i
$$
denotes the moment-generating function of $f_i$, the pdf of the
jump-sizes of $X$ in state $i$.

\proof{\it of Proposition \ref{EMM}} (i) Let $(X,Z)$ be of the
form \eqref{eq:Xt}, with corresponding characteristic matrix $K$
and suppose that $g$ is a $E^0$-row vector. Asmussen and Kella
\cite{AK} have shown that
\begin{equation}\label{eq:AKmart}
\te{b X_t - ct}g({Z_t}) - \te{bX_0}g({Z_0}) - \I_0^t \te{b X_u -
cu}g({Z_u})\td u( K[b] - c I),
\end{equation}
is a row vector of martingales for $c\ge0$ and $b$ such that the
diagonal elements of $K[b]$ are finite. Choosing in
(\ref{eq:AKmart}) the process $X$ to be equal to $X_{a}$, $b=1$,
$c=\l$ and $g=h$, it follows that $\te{X_{a}(t) - \l t}h_(Z_t) -
h_(Z_0)$ is a zero mean martingale. As $h$ is positive, the
process $L$ in (\ref{eq:martL}) is thus a positive mean one
martingale. The proof of (ii) can be found in the next section.
\exit

\subsection{Change of measure}
\begin{Prop}\label{prop:com}
Under $\P^*$, the process $X$ is still of the form \eqref{eq:Xt}
with $\s^*(i)=\s(i)$,
$$
\m^*(i) = \m(i) + \a_i\s_i^2 - \I_0^1y(1-\te{\a_i
y})\l^{(+)}_iF^{(+)}_i(\td y)
$$
and with $J^*_i$ compound Poisson processes with changed jump
rates
$$\l^{(+)*}_i = \l^{(+)}\H F^{(+)}[-\g]\q\text{and}\q
\l^{(-)*}_i = \l^{(-)}\H F^{(-)}[\g]
$$
and distributions of the positive and negative jumps of phase-type
with representations
$$(\a^{(+)*}_i,  T^{(+)*}_i) =
(\a^{(+,a_i)}_i,  T^{(+,a_i)}_i)\ \text{and}\ (\a^{(-)*}_i,
T^{(-)*}_i) = (\a^{(-,a_i)}_i,  T^{(-,a_i)}_i),
$$
respectively, where the parameters are transformed according to
\begin{eqnarray}\label{eq:paratilt}
(\a^{(+,\g)}, T^{(+,\g)}) &=& (\a^{(+)}\Delta_+/\H F^{(+)}[-\g],
\Delta^{-1}_{+} T^{(+)}\Delta_{+} + \g I),\\%
\label{eq:paratilt2} (\a^{(-,\g)}, T^{(-,\g)}) &=&
(\a^{(-)}\Delta_{-}/\H F^{(-)}[\g], \Delta^{-1}_{-} T^{(-)}
\Delta_{-} - \g I),
\end{eqnarray}
where $\Delta_{+}$ and $\Delta_-$ are the diagonal matrices with
respectively $(k_+)_i$ and $(k_-)_i$ on the diagonal such that  $
k_+ = (-\g I- T^{(+)})^{-1} t^{(+)}$, $ k_- = (\g I- T^{(-)})^{-1}
t^{(-)}$ and $ I$ is an identity matrix of appropriate size.
\end{Prop}
\begin{Prop}\label{prop:com2} Under $\P^*$, $Z$ has intensity matrix
$ G^*$ with elements
$$
g_{ij}^* = g_{ij}h(j)/h(i),\; i\neq j, \text{ and  $g_{ii}^* =
-\sum_{j\neq i}g_{ij}h(j)/h(i)$,}
$$
where $g_{ij}$ is the $ij$th element of $ G$;
\end{Prop}
\proof{\it of Propositions \ref{prop:com} and \ref{prop:com2}.} We
first show how to find the characteristic matrix of $X$ under
$\P^*$. Denote by $\WT X_{ a}$ the process $X_{ a} + bX$, where
$b$ is such that the elements of the characteristic matrix of $\WT
X_{ a}$ are finite, and let $f$ be a function that maps $E^0$ to
$\R$. Applying It\^{o}'s lemma to $\te{\WT X_{ a}(t) - \l t}h_{
a}(Z_t)f(Z_t)$ shows that
\begin{multline*}
\te{\WT X_{ a}(t) - \l t}h(Z_t)f(Z_t) - \te{\WT X_{
a}(0)}h(Z_0)f(Z_0)
- \I_0^t\l\te{\WT X_{ a}(u) - \l u}h(Z_u)f(Z_u)\td u\\
- \I_0^t\td u\; \te{\WT X_{ a}(u) - \l u} \times \\
\times \le[\sum_{i\in E^0}(\mbf 1_{\{Z_u=i\}}%
h(i)f(i)\k_i(a_i+b) +\sum_{j\neq i}g_{ij}(h(j)f(j)-h(i)f(i)))\ri],
\end{multline*}
is a $\P$-martingale, where we wrote $\l=\l_{ a}$ and $h=h_{ a}$.
Since
$$
\l_{ a}( h_{ a})_i = ( K_{ a}[1] h_{ a})_i = \sum_{j\neq
i}g_{ij}(h_{ a}(j) - h_{ a}(i)) + \k_i(a_i)h_{ a}(i),
$$
it follows from taking expectations and rearranging terms that, in
vector notation,
\begin{equation}\label{eq:charode}
\E_{0,i}[L_t\te{bX(t)}\mbf 1_{Z_t}] = \mbf 1_{i} + \I_0^t
\E_{0,i}[L_u\te{bX(u)}\mbf 1_{Z_u}] ( G^* + \D^*[b])\td u,
\end{equation}
where $ G^*$ is as in the statement of the proposition and $\D^*$
is the diagonal matrix with elements $\D^*_{ii} =
\k_i(a_i+b)-\k_i(a_i)$. Writing $ F^*_t[b]$ for the matrix with
elements $\E_{0,i}^*[\te{b X_t}\mbf 1_{Z_t=j}]$ and
differentiating (\ref{eq:charode}) with respect to $t$, we arrive
at the matrix differential equation
$$
 F^{*\prime}_t[b] = F^*_t[b] ( G^*+\D^*), \q  F^*_0[b]= I,
$$
where $\prime$ denotes the time derivative. Solving this system
shows that the characteristic matrix of $X$ under $\P^*$ is given
by $ K^*[b]= G^* + \D^*$. See \cite{Asm89} for a proof of
(\ref{eq:paratilt}) and (\ref{eq:paratilt2}); the rest of the
statements of (i) and (ii) directly follow from Proposition 3 in
\cite{PR}.\exit

\proof{\it of Proposition \ref{EMM}}(ii) From Proposition
\ref{prop:com} it follows that the $a_i$'s have been chosen in
such that, under $\P^*$, $X^i$ have cumulant-generating functions
satisfying $\k_i^{*}(1) =\k_i(a_i+1)-\k_i(a_i) = r_i$, so that the
characteristic matrix $K^*$ of $(X - \int_0^\cdot r(Z_s)\td s,Z)$
under $\P^*$ satisfies $K^*[1] =  G^*$ and $\mbf 1$ is an
eigenvector of $K^*[1]$ with eigenvalue $0$. Setting $g\equiv 1$
in (\ref{eq:AKmart}) and taking $a=0$, it thus follows that the
process $\te{X_t - \int_0^t r(Z_s)\td s}$ is a martingale under
$\P^*$. \exit

\subsection{Wiener-Hopf factorization}
\proof{\it of Theorem \ref{thm:symemb}}(ii) Now we turn to the
proof of the uniqueness of the Wiener-Hopf factorization. To this
end, let $( Z^+, G^+,  Z^-,  G^-)$ be a Wiener-Hopf factorization
and define the function $\WT f$ as in (\ref{eq:phi+-}), but
replacing $ \eta^+$ and $ Q^+$ by $ Z^+$ and ${ G^+}$
respectively. Since $( Z^+,  G^+)$ satisfies the first equation of
(\ref{systemnoiseWH3}), it follows from an application of
It\^{o}'s lemma, that $\WT f(Y_t, A_t)$ is a local martingale that
is bounded on $\{t\leq \t^+_\ell\}$, so that Doob's Optional
Stopping Theorem implies that
\begin{align}\nn
\WT{f}(j,x) &= \E_{x,j}[\WT f(Y_{t\wedge \t^+_\ell}, A_{t\wedge\t^+_\ell})]\\
&= \E_{x,j}[\WT f(\WT Y^+_{\ell},
A_{\t^+_\ell})1_{(\t^+_\ell<\i)}] + \lim_{t\to\i}\E_{x,j}[\WT
f(Y_{t}, A_{t})1_{(\t^+_\ell=\i)}]. \label{optionals}
\end{align}
By the definition of $\WT f$ and the absence of positive jumps of
$A$, the first expectation in (\ref{optionals}) is equal to
$f(j,x)$. Note that the second term in (\ref{optionals}) is zero
if $ Q$ is transient or $ Q$ is recurrent and $\sup_t A_t=+\i$.
Indeed, in the latter case, $\tau^+_\ell$ is finite a.s. whereas
in the former case $\P_{x,i}(Y_t\in E)$ converges to zero. Thus
$f=\WT f$ for all $ h$ and we deduce that $ G^+= Q^+$ and $
Z^+=\eta^+$. Similarly, one can show that $ G^-= Q^-$ and $
Z^-=\eta^-$ and the uniqueness is proved. \exit

\proof{\it of Theorem \ref{thm:symemb}}(iii) Assume that $ Q$ is
recurrent but $A_t\to-\i$. As $ Q^+$ inherits the irreducibility
property of $ Q$, it follows from the Perron-Frobenius theorem
that the matrix $ Q^+$ has a probability vector $\m$ as
left-eigenvector with its largest eigenvalue. Since the quadruple
$(\eta^+,  Q^+, \eta^-, Q^-)$ satisfies (\ref{systemnoiseWH3}), it
is straightforward to check that this remains the case if we
replace $(\eta^+,  Q^+)$ by $(\eta^+( I - \mathbf 1 \mu) + \mu,
Q^+( I -  \mbf 1\mu))$. We are left to show that these are the
only two factorizations of $(A, Y)$. As in the proof of Theorem
\ref{thm:symemb}(ii), it follows that any factorization quadruple
of $(A,Y)$ must contain $\eta^-$ and $ Q^-$.
Letting $(\eta^+, G^+)$ and $\WT f(j,x)$ be as in the proof of
Theorem \ref{thm:symemb}, we distinguish between the cases  that $
G$ is recurrent or transient. In the latter case $\WT f(j,x)$
tends to zero if $x\to-\i$ and we deduce from (\ref{optionals})
that $f=\WT f$ and thus $ G= Q^+$ and $ Z^+ =  \eta^+$. In the
former case, we note that, as $ G^+$ inherits the irreducibility
property of $ Q$, it has a unique invariant distribution $\nu$
given by the left-eigenvector of $ G$ with eigenvalue $0$. Thus
$\WT f(j, x)$ converges to $e_j'\mbf 1\nu h = \nu h$ as $x\to\i$.
The right-hand side of (\ref{optionals}) is thus equal to
\begin{align}\nn
\WT f(j,x) &= f(j,x) + \P_{x,j}(\t^+_{k}=\i) \nu  h\\
 &= f(j,x) + \le(1-e_j' W^+
\exp( Q^+(k-x))\mbf 1\ri)\nu h. \label{diffmu}
\end{align}
By differentiation of (\ref{diffmu}) with respect to $x$, we
deduce that $ G =  Q^+( I - \mbf 1\nu)$. In particular, it follows
that $\nu$ is a left-eigenvector of $ Q^+$. Since the
Perron-Frobenius eigenvector is the unique eigenvector with the
largest eigenvalue, it follows that $\mu=\nu$ and then also that $
Z^+ = \eta^+( I - \mathbf 1 \mu) + \mu$, which completes the
proof. \exit

\subsection{First-passage under state-dependent levels}
\proof{\it of Theorem \ref{thm:fipakz}} For brevity of notation we
will drop the subscript and write $v$ for $v_{b,k}$. Appealing to
the strong Markov property, it follows that, for $x>k_1$,
\begin{eqnarray*}
v(x,i) &=& \E_{x,i}[v(k_1,Y_{\t^-})1_{(\t^- < \zeta)}] =
\E_{x,i}[h^-_1(Y_{\t^-})1_{(\t^- < \zeta)}],
\end{eqnarray*}
where $\t^-=\t^-_{k_1}$, and  for $k_j < x < k_{j-1}$,
\begin{eqnarray*}
v(x,i) &=& \E_{x,i}\le[v(k_{j-1}, Y_\t)1_{(\t<\zeta, A_\t = k_{j-1})}\ri]\\
&+& \E_{x,i}\le[v(k_{j}, Y_\t)1_{(\t<\zeta, A_\t = k_{j})}\ri] +
\E_{x,i}\le[v(A_{\zeta-}, Y_{\zeta-})1_{(\zeta<\t)}\ri].
\end{eqnarray*}
Invoking results from Proposition \ref{prop:twoexit} yields that
\eqref{eq:vvv} is valid for some vectors $h^-_{j}, h^+_{j}$ and $
h^\dagger_{j}$. To finish the proof we have to show that the
stated form of these vectors is correct. We start with noting
that, by the structure of the process $(A,Y)$,
$$v(k_j,j) = \te{bk_j} \quad\text{
and}\quad v(k_j,\ell) = \te{bk_j} e_\ell'(sI - T_j^-)^{-1}t^-_j
\quad\text{for $\ell\in E_j^-$}.
$$
Furthermore, we claim that $v(\cdot, i)$ is continuous. Indeed,
from the Markov property it follows that for $\ell\in E_m^-$,
$m\in E^0$
\begin{equation}
v(z,\ell) = \int_0^\i v(z+y,m) e_\ell'\te{T_m^-y}t^-_m \td y,
\end{equation}
so that, in particular, it holds that $v(\cdot,\ell)$ is
continuously differentiable on $(k_m,\i)$. Similarly, it follows
that $v(\cdot,\ell)\in C^1(k_m,\i)$ for $\ell\in E_m^+$. The
continuity of $v(\cdot,\ell)$ for $\ell\in E^0$ follows directly
from its definition. As a consequence it follows that the equation
\eqref{eq:VD} holds true. Let $\ell>j, \ell\in E^0$ and consider
$v(x,\ell)$ for $x\in[k_j-\e,k_j + \e]$. By a Feynman-Kac argument
it follows that, on $[k_j-\e,k_j + \e]$ for $\e>0$ small enough
(such that $k_j-\e > k_\ell$), $v(\cdot,\ell)$ is equal to the
unique $C^2$ solution of the ODE
$$
\frac{\s^2(\ell)}{2} f'' + \m(\ell) f' - c(\ell) f = g, \quad\quad
f(k_j\pm\e) = v(k_j\pm \e,\ell),
$$
for some continuous function $g$ and some constant $c(\ell)$. In
particular, $v(\cdot,\ell)$ is continuously differentiable at
$k_j$ and it follows that \eqref{eq:vC} holds true. \exit

\subsection{American put}
\proof{\it of Theorem \ref{thm:mainap}} The proof of this result
follows a standard approach for solving perpetual American option
pricing problems. As argued above the optimal stopping time must
be of the form \eqref{eq:Tk1k2}. Therefore, the value function is
given by $V_{k^*}$ for some vector $k^*\in (-\i,\log K)^N$. The
vector $k^*$ can subsequently be found by optimisation. At this
point we note that the condition \eqref{eq:coefc} implies that for
the embedding $s(i)^2/2 + m(i)  < -q_{ii}$ is satisfied for all
$i\in E$, so that we can apply Theorem \ref{thm:fipakz}. Since,
for fixed $(x,i)$, $k\mapsto V_k(x,i)$ is continuously
differentiable it follows that $k^*$ satisfies
\begin{equation}\label{eq:opt}
\le.\frac{\partial V_k}{\partial k_j}(\te{x}, i)\ri|_{k=k^*} = 0
\q\text{for all $(x,i)$, $j=1,\ldots, N$}.
\end{equation}
Consider next the finite difference $[V_k(\te{k_j+h},j) -
V_k(\te{k_j},j)]/h$ and note that it is equal to the sum
\begin{equation}
\frac{V_k(\te{k_j+h},j) - V_{k+h}(\te{k_j+h},j)}{h} +
\frac{V_{k+h}(\te{k_j+h},j) - V_k(\te{k_j},j)}{h}.
\end{equation}
Letting $h\downarrow 0$, it follows from \eqref{eq:opt}, that the
first term converges to zero, while the second term converges to
$-\te{k_j}$. Thus we see that the smooth fit equations
\eqref{eq:smoothfit} hold true. By a martingale argument it also
follows that $V_{k^*} = V^*$ for any solution $k^*\in (-\i,\log
K)^N$ of \eqref{eq:smoothfit}. \exit

\proof{\it of Lemma 1} Suppose first that $$\alpha \neq
[\mu_1+\sqrt{\mu_1^2+2(r_1+q_1)\sigma_1^2} ]/{\sigma_1^2}.$$ From
the definitions of $g(\theta), F_1(\theta)$, and $F_2(\theta)$, we
have that
\begin{equation*}
g(+\infty)=+\infty, \quad g(-\infty)=-\infty,\quad
g(0)=\alpha[(q_1+r_1)(q_2+r_2)-q_1q_2]>0.
\end{equation*}
Note that $F_1(\theta)$ has two different real roots $\theta_{0,1}
> 0>\theta_{0,2}$ with
$$\theta_{0,2}=-[\mu_1+\sqrt{\mu_1^2+2(r+q_1)\sigma_1^2}
]/{\sigma_1^2},$$ we then have $\theta_{0,2} \neq -\alpha$. Also,
$$g(\theta_{0,1})=-q_1q_2(\alpha+\theta_{0,1})<0,$$ because $q_1,q_2,
\alpha, \theta_{0,1} >0$. Therefore, we further have that
\begin {eqnarray*}
g(\theta_{0,2})&=&
  \begin{cases}
    -q_1q_2(\alpha+\theta_{0,2})<0, & \text{if} \q\theta_{0,2}>-\alpha, \\
   -q_1q_2(\alpha+\theta_{0,2})>0, & \text{if} \q\theta_{0,2}<-\alpha.
  \end{cases} \\
\lim_{\theta \rightarrow -\alpha} g(\theta)&=&
  \begin{cases}
    \lambda\alpha F_1(-\alpha;r)>0, & \text{if} \q\theta_{0,2}>-\alpha, \\
    \lambda\alpha F_1(-\alpha;r)<0, & \text{if} \q\theta_{0,2}<-\alpha.
  \end{cases}
\end {eqnarray*}
In view of the intermediate value theorem the proof of the first
assertion is complete. Since $Q^-$ is a generator matrix, it is
negative semi-definite and the final assertion follows.

If $$\a=[\mu_1+\sqrt{\mu_1^2+2(r_1+q_1)\sigma_1^2}
]/{\sigma_1^2},$$ $\th=-\alpha$ is a root. By a similar reasoning
applied to $h(\th) = g(\th)/(\th + \alpha)$ and $h(-\alpha)<0$ it
can be shown that $h$ has four distinct roots (two positive and
two negative ones). \exit }

\end{document}